\documentclass[
aps,
prd,
reprint,
superscriptaddress,
nofootinbib
]{revtex4-2}

\usepackage[utf8]{inputenc}
\usepackage[T1]{fontenc}

\usepackage{amsmath,amssymb,bm}
\usepackage{graphicx}
\usepackage{xcolor}
\usepackage{booktabs}
\usepackage{array}
\usepackage{multirow}
\usepackage{slashed}
\usepackage{dcolumn}
\usepackage{mathrsfs}
\usepackage{microtype}

\definecolor{linkblue}{RGB}{0,70,150}
\definecolor{citered}{RGB}{150,35,35}
\definecolor{urlpurple}{RGB}{100,40,150}

\usepackage[
colorlinks=true,
linkcolor=linkblue,
citecolor=citered,
urlcolor=urlpurple
]{hyperref}

\IfFileExists{orcidlink.sty}{
	\usepackage{orcidlink}
}{
	\newcommand{\orcidlink}[1]{%
		\href{https://orcid.org/#1}{\textsuperscript{\scriptsize ORCID}}}
}

\newcommand{\dd}{\mathrm{d}}

\begin{document}
	
	% ============================================================
	% Title
	% ============================================================
	\title{
		Information Geometry of DGLAP Evolution:
		From Geodesic Tests to Emergent Small-Circle Geometry
	}
	
	% ============================================================
	% Authors
	% ============================================================
	\author{Wei Kou\orcidlink{0000-0002-4152-2150}}
	\email{kouwei@impcas.ac.cn}
	
	\affiliation{
		Institute of Modern Physics,
		Chinese Academy of Sciences,
		Lanzhou 730000,
		Gansu Province,
		China
	}
	
	\affiliation{
		Southern Center for Nuclear Science Theory (SCNT),
		Institute of Modern Physics,
		Chinese Academy of Sciences,
		Huizhou 516000,
		Guangdong Province,
		China
	}
	
	\affiliation{
		School of Nuclear Science and Technology,
		University of Chinese Academy of Sciences,
		Beijing 100049,
		China
	}
	
	\affiliation{
		State Key Laboratory of Heavy Ion Science and Technology,
		Institute of Modern Physics,
		Chinese Academy of Sciences,
		Lanzhou 730000,
		China
	}

	\author{Xurong Chen}
	\email{xchen@impcas.ac.cn}
	
	\affiliation{
		Institute of Modern Physics,
		Chinese Academy of Sciences,
		Lanzhou 730000,
		Gansu Province,
		China
	}
	
	\affiliation{
		Southern Center for Nuclear Science Theory (SCNT),
		Institute of Modern Physics,
		Chinese Academy of Sciences,
		Huizhou 516000,
		Guangdong Province,
		China
	}
	
	\affiliation{
		School of Nuclear Science and Technology,
		University of Chinese Academy of Sciences,
		Beijing 100049,
		China
	}
	
	\affiliation{
		State Key Laboratory of Heavy Ion Science and Technology,
		Institute of Modern Physics,
		Chinese Academy of Sciences,
		Lanzhou 730000,
		China
	}
	
	% ============================================================
	% Abstract
	% ============================================================

	\begin{abstract}
		We investigate the geometric structure of leading-order (LO)
		nonsinglet DGLAP evolution on the Fisher--Rao manifold of
		normalized valence distributions, asking whether the QCD-selected
		trajectory is geodesic and how its departure from geodesic motion
		is organized. Through the isometric embedding
		$\Phi=2\sqrt{p}$, the evolution is represented as a curve on
		a Hilbert sphere of radius two, with its velocity and acceleration
		determined directly by the DGLAP operator. For a GRV98-based
		total-valence trajectory over
		$2\leq Q^2\leq 2000~{\rm GeV}^2$, we find a nonzero but
		weakly scale-dependent geodesic curvature. An arc-length-weighted
		principal-component analysis shows that two affine directions
		capture more than $99.999\%$ of the trajectory variance. The
		corresponding affine plane intersects the Fisher sphere in a
		small circle, whose predicted geodesic curvature closely agrees
		with that reconstructed from the DGLAP-generated acceleration.
		An $x$-resolved analysis further reveals that Fisher motion
		is governed by relative PDF reshaping rather than probability
		weight alone, with different momentum-fraction regions
		contributing to motion along the trajectory and its transverse
		bending. The Mellin representation provides a complementary
		perspective on the effective dimensional reduction through the
		collective evolution of the spectral hierarchy. These results
		establish a quantitative information-geometric characterization
		of the reference LO DGLAP trajectory and reveal an organized,
		near-small-circle structure over the finite evolution interval.
	\end{abstract}

	\maketitle

	% ============================================================
	% I. Introduction
	% ============================================================
	\section{Introduction}
	\label{sec:introduction}
	
	Parton distribution functions (PDFs) encode the longitudinal momentum
	structure of hadrons, and their resolution-scale dependence is governed
	by the Dokshitzer--Gribov--Lipatov--Altarelli--Parisi (DGLAP) equations
	~\cite{GribovLipatov:1972a,GribovLipatov:1972b,Dokshitzer:1977,
		AltarelliParisi:1977}. This evolution is conventionally described through
	the changing PDFs and their Mellin moments
	~\cite{CollinsSoper:1982,CollinsSoperSterman:1989,EthierNocera:2020}.
	For a conserved partonic sector, however, a positive distribution can
	also be normalized to a probability density, allowing its scale
	evolution to be viewed as a trajectory on a statistical manifold.
	This raises a complementary question: what geometric path does QCD
	evolution select in the space of normalized distributions? The
	Fisher--Rao metric provides a natural measure of the statistical
	distinguishability between neighboring states
	~\cite{Fisher:1922,Rao:1945,AmariNagaoka:2000,Amari:2016}.
	
	Information-theoretic approaches to partonic structure have explored
	connections among small-$x$ dynamics, entanglement entropy, and
	final-state hadron production, including comparisons with DIS
	phenomenology
	~\cite{KharzeevLevin:2017,KharzeevLevin:2021,Kutak:2011,
		TuKharzeevUllrich:2020,H1:2021,HentschinskiKutak:2022,
		HentschinskiKutakStraka:2022,HentschinskiKharzeevKutakTu:2023,
		HentschinskiKharzeevKutakTu:2024}. These studies motivate broader
	information-theoretic questions in QCD, but their entropy observables
	are distinct from the statistical geometry of the evolving PDF
	considered here.
	
	Geometric descriptions of renormalization-group flows and
	information-theoretic coarse graining provide another related context
	~\cite{Dolan:1994,Dolan:1997,Dolan:1998,DolanLewis:1999,Kar:2001,
		Apenko:2012,BenyOsborne:2015PRA,BenyOsborne:2015NJP,
		Maity:2015,Erdmenger:2020,CotlerRezchikov:2023,Floerchinger:2024}.
	Such constructions often involve spaces of couplings, theories, or
	coarse-grained states. Here the manifold instead consists of
	normalized parton distributions. The DGLAP kernel determines their
	evolution, while the Fisher--Rao metric characterizes the geometry of
	the resulting path. The dynamical law and the statistical metric thus
	serve distinct, complementary roles.
	
	The leading-order (LO) nonsinglet valence sector offers a direct
	setting for this analysis. Its evolution does not involve
	singlet--gluon mixing, and valence-number conservation permits the
	positive valence distributions considered here to be normalized as
	probability densities $p(x,Q^2)$. The square-root embedding
	$\Phi=2\sqrt{p}$ maps these states isometrically into the positive
	sector of a Hilbert sphere of radius two
	~\cite{AmariNagaoka:2000,Ay:2017,BruverisMichor:2019}.
	DGLAP evolution therefore defines a curve whose length, curvature,
	and effective dimensionality can be studied in a common geometric
	framework.
	
	Although DGLAP evolution is governed by a first-order equation, the
	geodesic character of its trajectory is determined by the acceleration
	induced by that evolution. We obtain the first and second derivatives
	of the Fisher state directly from the LO evolution operator. Removing
	the sphere-normal component and the component parallel to the
	instantaneous velocity isolates the transverse geodesic acceleration,
	whose norm relative to the squared Fisher speed defines the
	parameterization-invariant geodesic curvature. This construction tests
	the geodesicity of the QCD-selected path without introducing a new
	evolution equation.
	
	We apply this framework to a GRV98-based total-valence distribution
	evolved over $2\leq Q^2\leq2000~{\rm GeV}^2$. The resulting Fisher
	trajectory is nongeodesic, yet its geodesic curvature varies only
	weakly across the analyzed interval. Arc-length-weighted principal-component analysis identifies a
	dominant two-dimensional affine
	sector, whose displaced plane defines a small-circle reference on the
	Fisher sphere. The curvature obtained directly from DGLAP-generated
	acceleration agrees closely with the geometric prediction of this
	construction.
	
	An $x$-resolved analysis connects the trajectory geometry to the
	relative reshaping of the normalized valence distribution: the regions
	contributing most strongly to statistical displacement need not be
	those dominating its transverse bending. The Mellin representation
	provides a complementary perspective on the collective evolution
	underlying the effective low-dimensional trajectory. Together, these
	results identify a nongeodesic but near-small-circle geometry for the
	reference LO nonsinglet evolution over a finite scale interval.
	
	% ============================================================
	% II. Information-geometric formulation
	% ============================================================
	
	\section{Information-Geometric Formulation of LO DGLAP Evolution}
	\label{sec:formalism}
	
	We formulate LO nonsinglet DGLAP evolution as a trajectory on the
	Fisher--Rao manifold of normalized valence distributions. The
	statistical geometry is fixed by probability normalization, whereas
	the physical path is selected by the DGLAP kernel. We first construct
	the Fisher embedding and then derive the velocity, acceleration, and
	geodesic curvature generated by the QCD evolution. Details concerning
	reparameterization invariance and numerical implementation are given in
	the appendices.

	\subsection{Normalized valence distribution and Fisher sphere}
	\label{subsec:probability-manifold}
	
	We consider a nonsinglet valence distribution
	\begin{equation}
		q_v(x,Q^2)
		=
		q(x,Q^2)-\bar q(x,Q^2),
		\label{eq:valence-def}
	\end{equation}
	with conserved valence number
	\begin{equation}
		N_v
		=
		\int_0^1 \dd x\,q_v(x,Q^2).
		\label{eq:valence-number}
	\end{equation}
	For nonsinglet DGLAP evolution, $N_v$ is independent of $Q^2$ because
	the first Mellin moment of the nonsinglet splitting kernel vanishes
	~\cite{GribovLipatov:1972a,GribovLipatov:1972b,Dokshitzer:1977,
		AltarelliParisi:1977,CurciFurmanskiPetronzio:1980}. We therefore define
	the normalized density
	\begin{equation}
		p(x,Q^2)
		=
		\frac{q_v(x,Q^2)}{N_v},
		\qquad
		\int_0^1 \dd x\,p(x,Q^2)=1.
		\label{eq:normalized-pdf}
	\end{equation}
	For the positive valence distributions considered here, $p(x,Q^2)$ may
	be treated as a probability density on $0<x<1$. Because $N_v$ is scale
	independent, the normalization removes only a conserved overall factor
	and leaves the shape evolution unchanged. The scale dependence
	therefore defines a trajectory
	\begin{equation}
		Q^2
		\longmapsto
		p(\,\cdot\,,Q^2)
		\label{eq:statistical-trajectory}
	\end{equation}
	on the infinite-dimensional manifold of normalized positive densities.
	
	An infinitesimal tangent deformation satisfies
	\begin{equation}
		\int_0^1 \dd x\,\delta p(x,Q^2)=0,
		\label{eq:tangent-normalization}
	\end{equation}
	and the Fisher--Rao line element is
	\begin{equation}
		\dd s_{\rm FR}^2
		=
		\int_0^1 \dd x\,
		\frac{\left[\delta p(x,Q^2)\right]^2}{p(x,Q^2)}.
		\label{eq:fisher-rao-line-element}
	\end{equation}
	This metric provides the intrinsic local measure of statistical
	distinguishability on the probability manifold
	~\cite{Fisher:1922,Rao:1945,AmariNagaoka:2000,Ay:2017,
		BruverisMichor:2019}.
	
	The spherical form of the metric follows directly from the
	square-root embedding
	\begin{equation}
		\Phi(x,Q^2)
		=
		2\sqrt{p(x,Q^2)}.
		\label{eq:fisher-embedding}
	\end{equation}
	Normalization implies
	\begin{equation}
		\int_0^1 \dd x\,\Phi^2(x,Q^2)=4,
		\label{eq:fisher-sphere-radius}
	\end{equation}
	while
	\begin{equation}
		\delta\Phi(x,Q^2)
		=
		\frac{\delta p(x,Q^2)}{\sqrt{p(x,Q^2)}}
	\end{equation}
	gives
	\begin{equation}
		\dd s_{\rm FR}^2
		=
		\int_0^1 \dd x\,
		\left[\delta\Phi(x,Q^2)\right]^2.
		\label{eq:fisher-euclidean-metric}
	\end{equation}
	Thus $\Phi=2\sqrt p$ is an isometric embedding of the Fisher--Rao
	manifold into the positive sector of a Hilbert sphere~\cite{AmariNagaoka:2000,Ay:2017,BruverisMichor:2019,Wootters:1981} with radius
	\begin{equation}
		R_{\rm FR}=2.
		\label{eq:fisher-radius}
	\end{equation}
	
	For two normalized states $p_1(x)$ and $p_2(x)$, their embedded inner
	product is determined by the Bhattacharyya overlap~\cite{Bhattacharyya:1943},
	\begin{equation}
		\mathcal{B}(p_1,p_2)
		=
		\int_0^1 \dd x\,\sqrt{p_1(x)p_2(x)},
		\ 
		\langle\Phi_1,\Phi_2\rangle
		=
		4\,\mathcal{B}(p_1,p_2).
		\label{eq:bhattacharyya-overlap}
	\end{equation}
	Their angular separation on the Fisher sphere is therefore
	$\Theta_{12}=\arccos\mathcal{B}(p_1,p_2)$, and the corresponding
	Fisher--Rao geodesic distance is
	\begin{equation}
		d_{\rm FR}(p_1,p_2)
		=
		2\arccos
		\left[
		\int_0^1 \dd x\,
		\sqrt{p_1(x)p_2(x)}
		\right].
		\label{eq:fisher-geodesic-distance}
	\end{equation}
	This distance is fixed entirely by the endpoint distributions. DGLAP
	evolution, in contrast, selects a specific physical path between those
	endpoints. The distinction between the metric shortest path and the
	QCD-selected trajectory will be central to the analysis below.
	
	\subsection{LO DGLAP flow on the statistical manifold}
	\label{subsec:dglap-flow}
	
	We now introduce the QCD dynamics that selects the physical trajectory
	on the Fisher--Rao manifold. Using
	\begin{equation}
		t=\ln\frac{Q^2}{Q_0^2},
		\label{eq:t-def}
	\end{equation}
	the LO nonsinglet DGLAP equation for the normalized valence density is
	\begin{equation}
		\frac{\partial p(x,t)}{\partial t}
		=
		a(t)\,
		\left[P_{qq}^{(0)}\otimes p\right](x,t),
		\qquad
		a(t)=\frac{\alpha_s(Q^2)}{2\pi},
		\label{eq:dglap-t}
	\end{equation}
	where
	\begin{equation}
		\left[P\otimes f\right](x)
		=
		\int_x^1\frac{\dd z}{z}\,
		P(z)\,
		f\!\left(\frac{x}{z}\right).
		\label{eq:convolution}
	\end{equation}
	Because the first Mellin moment of $P_{qq}^{(0)}$ vanishes, this flow
	preserves the normalization of $p$ and therefore remains tangent to the
	statistical manifold.
	
	For the geometric analysis it is convenient to absorb the running
	coupling into the evolution parameter,
	\begin{equation}
		\tau(Q^2)
		=
		\int_{Q_0^2}^{Q^2}
		\frac{\dd\mu^2}{\mu^2}\,
		\frac{\alpha_s(\mu^2)}{2\pi},
		\qquad
		\tau(Q_0^2)=0,
		\label{eq:tau-def}
	\end{equation}
	so that
	\begin{equation}
		\frac{\dd\tau}{\dd t}=a(t).
		\label{eq:tau-t-relation}
	\end{equation}
	The LO evolution then takes the fixed-operator form
	\begin{equation}
		\frac{\partial p(x,\tau)}{\partial\tau}
		=
		\mathcal{L}p(x,\tau),
		\qquad
		\mathcal{L}f
		\equiv
		P_{qq}^{(0)}\otimes f .
		\label{eq:dglap-tau}
	\end{equation}
	At LO, $\mathcal{L}$ is independent of $\tau$. The running coupling
	therefore determines the mapping between the physical scale $Q^2$ and
	the evolution parameter $\tau$, whereas the path in the space of
	normalized distributions is generated by the fixed operator
	$\mathcal{L}$.
	
	Under the Fisher embedding $\Phi=2\sqrt{p}$, the corresponding velocity
	is
	\begin{equation}
		\partial_\tau\Phi(x,\tau)
		=
		\frac{\partial_\tau p(x,\tau)}{\sqrt{p(x,\tau)}}
		=
		\frac{\mathcal{L}p(x,\tau)}{\sqrt{p(x,\tau)}} .
		\label{eq:fisher-velocity}
	\end{equation}
	Normalization conservation gives
	\begin{equation}
		\left\langle
		\Phi,\partial_\tau\Phi
		\right\rangle
		=
		2\int_0^1\dd x\,\partial_\tau p(x,\tau)
		=
		0,
		\label{eq:fisher-tangent}
	\end{equation}
	confirming that the DGLAP-induced velocity is tangent to the Fisher
	sphere.
	
	Its norm defines the Fisher--Rao speed in the $\tau$ parametrization,
	\begin{equation}
		v_{\rm FR}^{(\tau)}
		=
		\left\|
		\partial_\tau\Phi
		\right\|
		=
		\left[
		\int_0^1\dd x\,
		\frac{\left[\mathcal{L}p(x,\tau)\right]^2}
		{p(x,\tau)}
		\right]^{1/2}.
		\label{eq:fisher-speed-tau}
	\end{equation}
	The physical path length accumulated between two scales is then
	\begin{equation}
		L_{\rm DGLAP}
		=
		\int_{\tau_1}^{\tau_2}
		\dd\tau\,
		v_{\rm FR}^{(\tau)} .
		\label{eq:dglap-length}
	\end{equation}
	Although the instantaneous speed depends on the parameter used to
	traverse the trajectory,
	\begin{equation}
		v_{\rm FR}^{(t)}
		=
		a(t)\,v_{\rm FR}^{(\tau)},
		\label{eq:speed-transform}
	\end{equation}
	the accumulated Fisher--Rao length is invariant,
	\begin{equation}
		v_{\rm FR}^{(t)}\,\dd t
		=
		v_{\rm FR}^{(\tau)}\,\dd\tau
		=
		\dd s_{\rm FR}.
		\label{eq:length-invariance-local}
	\end{equation}
	The explicit reparameterization properties of the curvature are derived
	in Appendix~\ref{app:parameterization}.

	\subsection{Acceleration and direct geodesic test}
	\label{subsec:geodesic-test}
	
	Geodesicity is determined by the acceleration of a curve rather than by
	its velocity alone. Differentiating the fixed-operator evolution in
	Eq.~\eqref{eq:dglap-tau}, and using the $\tau$ independence of
	$\mathcal{L}$ at LO, gives
	\begin{equation}
		\partial_\tau^2 p
		=
		\mathcal{L}^2p.
		\label{eq:second-tau-derivative}
	\end{equation}
	The second derivative therefore involves only repeated action of the
	same LO evolution operator and does not introduce a higher-order
	splitting kernel.
	
	For the Fisher embedding $\Phi=2\sqrt{p}$, the corresponding
	acceleration is
	\begin{equation}
		\partial_\tau^2\Phi
		=
		\frac{\mathcal{L}^2p}{\sqrt{p}}
		-
		\frac{\left(\mathcal{L}p\right)^2}
		{2p^{3/2}} .
		\label{eq:fisher-acceleration}
	\end{equation}
	The second term originates from the nonlinear square-root embedding and
	is therefore geometric in origin even though the evolution equation
	for $p$ is linear.
	
	The spherical constraint
	\begin{equation}
		\langle\Phi,\Phi\rangle
		=
		R_{\rm FR}^2
		\label{eq:sphere-constraint}
	\end{equation}
	implies
	\begin{equation}
		\left\langle
		\Phi,\partial_\tau\Phi
		\right\rangle
		=
		0,
		\qquad
		\left\langle
		\Phi,\partial_\tau^2\Phi
		\right\rangle
		=
		-\left[v_{\rm FR}^{(\tau)}\right]^2 .
		\label{eq:sphere-derivative-identities}
	\end{equation}
	Hence the radial component of the acceleration is fixed entirely by the
	sphere geometry,
	\begin{equation}
		\left(\partial_\tau^2\Phi\right)_{\rm rad}
		=
		-
		\frac{\left[v_{\rm FR}^{(\tau)}\right]^2}
		{R_{\rm FR}^2}\,
		\Phi .
		\label{eq:radial-acceleration}
	\end{equation}
	Removing this radial contribution leaves the acceleration tangent to
	the Fisher sphere,
	\begin{equation}
		A_{\rm tan}^{(\tau)}
		=
		\partial_\tau^2\Phi
		+
		\frac{\left[v_{\rm FR}^{(\tau)}\right]^2}
		{R_{\rm FR}^2}\,
		\Phi .
		\label{eq:tangent-acceleration}
	\end{equation}
	
	The parameter $\tau$ is not generally an affine parameter along the
	curve, so $A_{\rm tan}^{(\tau)}$ may still contain a component parallel
	to the velocity. Such a component changes only the rate at which the
	trajectory is traversed and does not bend the path. Removing it defines
	the transverse geodesic residual,
	\begin{equation}
		\mathcal{R}_{\rm geo}^{(\tau)}
		=
		\partial_\tau^2\Phi
		+
		\frac{\left[v_{\rm FR}^{(\tau)}\right]^2}
		{R_{\rm FR}^2}\,
		\Phi
		-
		\frac{
			\left\langle
			\partial_\tau\Phi,\partial_\tau^2\Phi
			\right\rangle
		}{
			\left[v_{\rm FR}^{(\tau)}\right]^2
		}
		\partial_\tau\Phi .
		\label{eq:geodesic-residual}
	\end{equation}
	By construction,
	\begin{equation}
		\left\langle
		\Phi,\mathcal{R}_{\rm geo}^{(\tau)}
		\right\rangle
		=
		0,
		\qquad
		\left\langle
		\partial_\tau\Phi,\mathcal{R}_{\rm geo}^{(\tau)}
		\right\rangle
		=
		0,
		\label{eq:geodesic-residual-orthogonality}
	\end{equation}
	so that $\mathcal{R}_{\rm geo}^{(\tau)}$ measures only the transverse
	departure of the physical trajectory from Fisher--Rao geodesic motion.
	
	The intrinsic geodesic curvature is therefore
	\begin{equation}
		k_{\rm FR}
		=
		\frac{
			\left\|
			\mathcal{R}_{\rm geo}^{(\tau)}
			\right\|
		}{
			\left[v_{\rm FR}^{(\tau)}\right]^2
		}.
		\label{eq:geodesic-curvature}
	\end{equation}
	A Fisher--Rao geodesic satisfies
	\begin{equation}
		k_{\rm FR}=0.
		\label{eq:geodesic-condition}
	\end{equation}
	Although the velocity and acceleration separately depend on the
	parameter used to traverse the curve, the combination in
	Eq.~\eqref{eq:geodesic-curvature} is intrinsic. Its invariance under a
	monotonic reparameterization is shown explicitly in
	Appendix~\ref{app:parameterization}.
	
	\subsection{Great-circle and small-circle geometry}
	\label{subsec:small-circle-geometry}
	
	The local geodesic test can be connected to a simple global geometry.
	Suppose that the Fisher trajectory is confined, exactly or
	approximately, to a two-dimensional affine plane
	\begin{equation}
		\mathcal{P}
		=
		\left\{
		C+u_1 e_1+u_2 e_2
		\right\},
		\label{eq:affine-plane}
	\end{equation}
	where $e_1$ and $e_2$ are orthonormal directions within the plane,
	\begin{equation}
		\langle e_a,e_b\rangle=\delta_{ab},
		\qquad
		\langle C,e_a\rangle=0.
		\label{eq:affine-plane-orthogonality}
	\end{equation}
	The vector $C$ therefore points from the center of the Fisher sphere to
	the closest point of the affine plane, and its norm
	\begin{equation}
		h=\|C\|
		\label{eq:plane-offset}
	\end{equation}
	defines the plane offset.
	
	Because the Fisher sphere has radius $R_{\rm FR}$, the intersection of
	$\mathcal{P}$ with the sphere satisfies
	\begin{equation}
		u_1^2+u_2^2
		=
		R_{\rm FR}^2-h^2 .
		\label{eq:sphere-plane-intersection}
	\end{equation}
	The intersection is therefore a circle with radius
	\begin{equation}
		r
		=
		\sqrt{R_{\rm FR}^2-h^2}.
		\label{eq:small-circle-radius}
	\end{equation}
	If $h=0$, the affine plane passes through the sphere center and the
	intersection is a great circle, corresponding to a Fisher--Rao
	geodesic. For $h\neq0$, the intersection is a small circle.
	
	For an exact small circle, the intrinsic geodesic curvature is constant
	and is fixed entirely by the sphere radius and plane offset,
	\begin{equation}
		k_{\rm sc}
		=
		\frac{|h|}{R_{\rm FR}\,r}.
		\label{eq:small-circle-curvature}
	\end{equation}
	Thus the great-circle limit $h\rightarrow0$ gives
	$k_{\rm sc}\rightarrow0$, as required for geodesic motion.
	
	If the trajectory spans an angular interval $\Delta\varphi$ within the
	affine plane, the corresponding circular arc length is
	\begin{equation}
		L_{\rm sc}
		=
		r\,\Delta\varphi .
		\label{eq:small-circle-length}
	\end{equation}
	The quantities $r$, $k_{\rm sc}$, and $L_{\rm sc}$ are therefore
	predictions of the affine-plane geometry rather than independent fit
	parameters. In the numerical analysis below, they will be compared
	directly with the radius, geodesic curvature, and accumulated
	Fisher--Rao length of the DGLAP trajectory.

	% ============================================================
	% III. Numerical evolution and geometric observables
	% ============================================================
	\section{Numerical Evolution and Geometric Observables}
	\label{sec:numerics}
	
	We now construct a reference LO nonsinglet trajectory in $x$ space and
	evaluate the Fisher--Rao observables introduced in Sec.~\ref{sec:formalism}.
	The underlying DGLAP evolution is validated independently before the
	geometric analysis is performed. Numerical discretization, conservation
	corrections, and detailed validation tests are collected in
	Appendix~\ref{app:numerics}.

	\subsection{Reference LO evolution}
	\label{subsec:reference-evolution}
	
	For the reference trajectory, we use the total valence combination
	\begin{equation}
		q_v^{\rm tot}(x,Q^2)
		=
		u_v(x,Q^2)+d_v(x,Q^2)
		\label{eq:total-valence}
	\end{equation}
	from the LO GRV98 parametrization
	~\cite{GluckReyaVogt:1998}. The evolution is initialized at the
	original GRV98 input scale,
	\begin{equation}
		Q_{\rm GRV}^2
		=
		0.26~{\rm GeV}^2,
		\label{eq:grv-input-scale}
	\end{equation}
	and first evolved to
	\begin{equation}
		Q_0^2
		=
		2~{\rm GeV}^2.
		\label{eq:reference-scale}
	\end{equation}
	This pre-evolution constructs the initial statistical state used in the
	information-geometric analysis. No geometric interpretation is assigned
	to the evolution below $Q_0^2$; the lower-scale evolution is used only
	to construct the GRV98-based reference state at
	$Q_0^2=2~{\rm GeV}^2$.
	
	At $Q_0^2$, the normalized density is defined by
	\begin{equation}
		p(x,Q_0^2)
		=
		\frac{
			u_v(x,Q_0^2)+d_v(x,Q_0^2)
		}{
			\displaystyle
			\int_0^1\dd x\,
			\left[
			u_v(x,Q_0^2)+d_v(x,Q_0^2)
			\right]
		}.
		\label{eq:reference-density}
	\end{equation}
	In the continuum valence normalization the denominator equals three.
	In the numerical calculation, however, the normalization is evaluated
	with the same quadrature used for the evolution, as described in
	Appendix~\ref{app:numerics}.
	
	The Fisher--Rao trajectory is studied over
	\begin{equation}
		2~{\rm GeV}^2
		\leq
		Q^2
		\leq
		2000~{\rm GeV}^2.
		\label{eq:analysis-range}
	\end{equation}
	Thus $Q_{\rm GRV}^2$ is the dynamical input scale of the GRV98
	parametrization, whereas $Q_0^2$ defines the starting point of the
	geometric trajectory.
	
	The reference evolution uses the LO nonsinglet splitting function
	\begin{equation}
		P_{qq}^{(0)}(z)
		=
		C_F
		\left[
		\frac{2}{(1-z)_+}
		-1-z
		+\frac{3}{2}\delta(1-z)
		\right],
		\label{eq:reference-splitting-kernel}
	\end{equation}
	where $C_F=4/3$. This is equivalent to the conventional
	LO Altarelli--Parisi form.
	The plus distribution is defined through
	\begin{equation}
		\int_0^1\dd z\,
		\frac{f(z)}{(1-z)_+}
		=
		\int_0^1\dd z\,
		\frac{f(z)-f(1)}{1-z}.
		\label{eq:plus-prescription}
	\end{equation}
	Consequently,
	\begin{equation}
		\int_0^1\dd z\,
		P_{qq}^{(0)}(z)
		=
		0,
		\label{eq:first-moment-zero}
	\end{equation}
	which expresses valence-number conservation.
	
	The running coupling is treated consistently at one-loop order,
	\begin{equation}
		\alpha_s(Q^2)
		=
		\frac{4\pi}
		{\beta_0\ln(Q^2/\Lambda_{\rm QCD}^2)},
		\qquad
		\beta_0
		=
		11-\frac{2}{3}n_f .
		\label{eq:one-loop-running}
	\end{equation}
	The flavor thresholds, the corresponding values of
	$\Lambda_{\rm QCD}$, and the numerical discretization are specified in
	Appendix~\ref{app:numerics}.
	
	The continuum evolution and its first two derivatives follow
	Eqs.~\eqref{eq:dglap-tau} and~\eqref{eq:second-tau-derivative}.
	In the numerical calculation, the velocity and acceleration are
	constructed from the corresponding discrete DGLAP convolution, with
	the valence-normalization constraint imposed on the grid as described
	in Appendix~\ref{app:discrete-evolution}. The geometric derivatives
	are generated from this discrete evolution operator rather than by
	finite differences of the sampled trajectory.
	
	\subsection{Validation of the reference trajectory}
	\label{subsec:numerical-validation}
	
	Before constructing the geometric observables, we verify the underlying
	$x$-space evolution against the analytic LO nonsinglet solution in
	Mellin space. Defining
	\begin{equation}
		M_N(Q^2)
		=
		\int_0^1\dd x\,
		x^{N-1}p(x,Q^2),
		\label{eq:mellin-moment}
	\end{equation}
	the LO evolution becomes diagonal,
	\begin{equation}
		\partial_\tau M_N
		=
		\gamma_N^{(0)}M_N,
		\qquad
		\gamma_N^{(0)}
		=
		\int_0^1\dd z\,
		z^{N-1}P_{qq}^{(0)}(z),
		\label{eq:mellin-evolution}
	\end{equation}
	with solution
	\begin{equation}
		M_N(\tau)
		=
		M_N(0)
		\exp\!\left[
		\gamma_N^{(0)}\tau
		\right].
		\label{eq:mellin-solution}
	\end{equation}
	The conserved first moment, $\gamma_1^{(0)}=0$, provides
	an additional normalization check.
	
	For the first five Mellin moments, the moments reconstructed from the
	numerical $x$-space evolution reproduce
	Eq.~\eqref{eq:mellin-solution} over the full analysis interval. The
	maximum relative discrepancy among the nontrivial moments is
	\begin{equation}
		\epsilon_{\rm Mellin}^{\max}
		\simeq
		3.5\times10^{-4}.
		\label{eq:mellin-error}
	\end{equation}
	After imposing the normalization constraint on the numerical grid,
	the maximum deviation from unit discrete normalization is
	\begin{equation}
		\max_\tau
		\left|
		\sum_{j=1}^{N_x}w_j p_j(\tau)-1
		\right|
		\simeq 2.2\times10^{-16}.
		\label{eq:normalization-error}
	\end{equation}
	The evolved density remains positive throughout the trajectory, so that
	the Fisher embedding $\Phi=2\sqrt{p}$ is well defined over the complete
	numerical domain.
	
	The local geodesic curvature is reconstructed from the
	operator-generated first and second derivatives of the Fisher state.
	As a complementary geometric check, we compare it with a finite-chord
	curvature estimator constructed from neighboring points of the
	trajectory. The two reconstructions agree at the few-parts-per-million
	level, with a maximum relative mismatch of
	\begin{equation}
		\max
		\left|
		\frac{k_{\rm chord}}{k_{\rm FR}}-1
		\right|
		\simeq
		3.3\times10^{-6}.
		\label{eq:curvature-validation}
	\end{equation}
	
	The Mellin-space comparison, discrete normalization check, and
	finite-chord reconstruction provide complementary validations of
	the reference evolution and its geometric implementation. The
	numerical prescriptions and diagnostics are collected in
	Appendix~\ref{app:numerics}.
	
	\subsection{Geometric observables}
	\label{subsec:geometric-observables}
	
	We characterize the Fisher--Rao trajectory at three complementary
	levels. First, principal-component analysis (PCA)
	~\cite{JolliffeCadima:2016}
	determines whether the evolution is effectively confined to a
	low-dimensional affine sector. Second, the DGLAP-generated acceleration
	gives the local geodesic curvature. Third, the sphere--plane geometry
	provides global predictions for the circle radius and accumulated path
	length. All inner products are evaluated in the Fisher representation,
	\begin{equation}
		\langle f,g\rangle
		=
		\int_0^1\dd x\,
		f(x)g(x).
		\label{eq:fisher-inner-product}
	\end{equation}
	
	For the sampled Fisher states
	$\Phi_i=\Phi(\cdot,Q_i^2)$, we assign normalized arc-length weights
	$\omega_i$ using the trapezoidal discretization of the accumulated
	Fisher--Rao length,
	\begin{equation}
		\sum_i\omega_i=1.
		\label{eq:arc-length-weight-normalization}
	\end{equation}
	The weighted mean state is
	\begin{equation}
		\bar\Phi
		=
		\sum_i\omega_i\Phi_i,
		\label{eq:weighted-fisher-mean}
	\end{equation}
	and the centered trajectory is
	\begin{equation}
		\delta\Phi_i
		=
		\Phi_i-\bar\Phi.
		\label{eq:centered-fisher-state}
	\end{equation}
	The weighted covariance operator,
	\begin{equation}
		\mathcal{C}
		=
		\sum_i
		\omega_i\,
		\delta\Phi_i\otimes\delta\Phi_i,
		\label{eq:weighted-covariance}
	\end{equation}
	defines orthonormal principal directions $e_a$ with eigenvalues
	$\lambda_a$.
	
	The fraction of the total trajectory variance captured by the first
	two directions is
	\begin{equation}
		f_2
		=
		\frac{
			\lambda_1+\lambda_2
		}{
			\displaystyle
			\sum_a\lambda_a
		}.
		\label{eq:pca-fraction}
	\end{equation}
	When $f_2$ is close to unity, the trajectory is effectively confined to
	the affine plane
	\begin{equation}
		\mathcal{P}_{\rm PCA}
		=
		\bar\Phi
		+
		\operatorname{span}
		\{e_1,e_2\}.
		\label{eq:pca-plane}
	\end{equation}
	
	Let $\Phi_i^{\rm proj}$ denote the orthogonal projection of $\Phi_i$
	onto this plane. We define the pointwise out-of-plane displacement
	\begin{equation}
		\delta_{\perp,i}
		=
		\left\|
		\Phi_i-\Phi_i^{\rm proj}
		\right\|,
		\label{eq:pointwise-plane-residual}
	\end{equation}
	and the arc-length-weighted RMS size of the full trajectory,
	\begin{equation}
		\sigma_{\rm traj}
		=
		\left[
		\sum_i\omega_i
		\left\|
		\Phi_i-\bar\Phi
		\right\|^2
		\right]^{1/2}.
		\label{eq:trajectory-rms-size}
	\end{equation}
	The normalized RMS departure from the dominant affine sector is
	\begin{equation}
		\epsilon_{\rm plane}
		=
		\frac{\left[\sum_i\omega_i\delta_{\perp,i}^2\right]^{1/2}}
		{\sigma_{\rm traj}}.
		\label{eq:plane-residual}
	\end{equation}
	For the two-component PCA used here,
	\begin{equation}
		\epsilon_{\rm plane}
		=
		\sqrt{1-f_2}.
		\label{eq:plane-residual-pca}
	\end{equation}
	
	Writing the projected trajectory as
	\begin{equation}
		\Phi_i^{\rm proj}
		=
		C+u_{1,i}e_1+u_{2,i}e_2,
		\label{eq:projected-trajectory}
	\end{equation}
	where $C$ is the point of $\mathcal{P}_{\rm PCA}$ closest to the
	Fisher-sphere center, the plane offset is
	\begin{equation}
		h=\|C\|.
		\label{eq:numerical-plane-offset}
	\end{equation}
	The sphere--plane geometry of
	Sec.~\ref{subsec:small-circle-geometry} then determines the
	reference circle radius and curvature through
	Eqs.~\eqref{eq:small-circle-radius}
	and~\eqref{eq:small-circle-curvature}.
	
	For each projected state we define the in-plane radius
	\begin{equation}
		\rho_i
		=
		\sqrt{
			u_{1,i}^2+u_{2,i}^2
		},
		\label{eq:projected-radius}
	\end{equation}
	and quantify the radial departure from the sphere--plane prediction by
	the arc-length-weighted residual
	\begin{equation}
		\epsilon_{\rm rad}
		=
		\left[
		\sum_i
		\omega_i
		\left(
		\frac{\rho_i-r}{r}
		\right)^2
		\right]^{1/2}.
		\label{eq:radial-residual}
	\end{equation}
	Thus $f_2$ and $\epsilon_{\rm plane}$ characterize the effective
	dimensional reduction, whereas $\epsilon_{\rm rad}$ tests circular
	closure within the dominant affine plane.
	
	The local bending of the physical trajectory is characterized by
	the geodesic curvature in Eq.~\eqref{eq:geodesic-curvature}.
	Its comparison with $k_{\rm sc}$ tests whether the local nongeodesic
	bending is consistent with the geometry inferred independently from the
	dominant affine plane.
	
	For comparison with a constant-curvature reference, we define the
	Fisher--Rao arc-length average
	\begin{equation}
		\left\langle k_{\rm FR}\right\rangle_s
		=
		\frac{
			\displaystyle
			\int \dd\tau\,
			v_{\rm FR}^{(\tau)}
			k_{\rm FR}(\tau)
		}{
			\displaystyle
			\int \dd\tau\,
			v_{\rm FR}^{(\tau)}
		}.
		\label{eq:arc-averaged-curvature}
	\end{equation}
	The corresponding arc-length-weighted standard deviation is
	\begin{equation}
		\sigma_{k,s}
		=
		\left[
		\frac{
			\displaystyle
			\int \dd\tau\,
			v_{\rm FR}^{(\tau)}
			\left(
			k_{\rm FR}(\tau)
			-
			\left\langle k_{\rm FR}\right\rangle_s
			\right)^2
		}{
			\displaystyle
			\int \dd\tau\,
			v_{\rm FR}^{(\tau)}
		}
		\right]^{1/2}.
		\label{eq:arc-curvature-standard-deviation}
	\end{equation}
	
	The global path geometry is characterized by the accumulated
	Fisher--Rao length
	\begin{equation}
		L_{\rm DGLAP}(Q^2)
		=
		\int_0^{\tau(Q^2)}
		\dd\tau\,
		v_{\rm FR}^{(\tau)},
		\label{eq:cumulative-dglap-length}
	\end{equation}
	and by the small-circle arc length defined in
	Eq.~\eqref{eq:small-circle-length}, where $\Delta\varphi$ is
	the angular displacement within the dominant affine plane. For the corresponding small-circle reconstruction, we define
	\begin{equation}
		\eta_G^{\rm sc}
		=
		\frac{d_{\rm FR}}{L_{\rm sc}}
		=
		\frac{d_{\rm FR}}{r\,\Delta\varphi}.
		\label{eq:small-circle-geodesic-ratio}
	\end{equation}
	This provides the small-circle reference used in the comparison with
	the physical DGLAP trajectory. We also define the geodesic length ratio
	\begin{equation}
		\eta_G
		=
		\frac{
			d_{\rm FR}
		}{
			L_{\rm DGLAP}
		},
		\qquad
		0<\eta_G\leq1,
		\label{eq:geodesic-length-ratio}
	\end{equation}
	where $d_{\rm FR}$ is the Fisher--Rao geodesic distance between the
	endpoints. A minimizing Fisher--Rao geodesic has $\eta_G=1$, while
	$\eta_G<1$ measures the excess length of the QCD-selected path
	relative to the metric shortest path.
	
	To resolve the Fisher motion in momentum fraction, we define
	\begin{equation}
		\left[
		v_{\rm FR}^{(\tau)}
		\right]^2
		=
		\int_0^1\dd x\,
		\mathcal{I}_\tau(x,\tau),
		\qquad
		\mathcal{I}_\tau(x,\tau)
		=
		\frac{
			\left[
			\partial_\tau p(x,\tau)
			\right]^2
		}{
			p(x,\tau)
		}.
		\label{eq:fisher-speed-density}
	\end{equation}
	Its contribution per logarithmic interval in $x$ is
	\begin{equation}
		\frac{
			\dd
			\left[
			v_{\rm FR}^{(\tau)}
			\right]^2
		}{
			\dd\ln x
		}
		=
		x\,\mathcal{I}_\tau(x,\tau).
		\label{eq:log-fisher-speed-density}
	\end{equation}
	For an interval $[x_a,x_b]$, the fractional contribution to the
	squared Fisher speed is
	\begin{equation}
		F_{[x_a,x_b]}(\tau)
		=
		\frac{
			\displaystyle
			\int_{x_a}^{x_b}\dd x\,
			\mathcal{I}_\tau(x,\tau)
		}{
			\left[
			v_{\rm FR}^{(\tau)}
			\right]^2
		}.
		\label{eq:fisher-speed-fraction}
	\end{equation}
	
	The transverse bending is resolved by the acceleration
	$\mathcal{A}_{\perp}^{(\tau)}\equiv
	\mathcal{R}_{\rm geo}^{(\tau)}$ (transverse to the Fisher
	trajectory, not to the hadron momentum), with
	\begin{equation}
		\frac{
			\dd
			\left\|
			\mathcal{A}_{\perp}^{(\tau)}
			\right\|^2
		}{
			\dd\ln x
		}
		=
		x
		\left[
		\mathcal{A}_{\perp}^{(\tau)}(x,\tau)
		\right]^2 .
		\label{eq:log-transverse-density}
	\end{equation}
	The corresponding regional fraction is
	\begin{equation}
		B_{[x_a,x_b]}(\tau)
		=
		\frac{
			\displaystyle
			\int_{x_a}^{x_b}\dd x\,
			\left[
			\mathcal{A}_{\perp}^{(\tau)}(x,\tau)
			\right]^2
		}{
			\displaystyle
			\int_0^1\dd x\,
			\left[
			\mathcal{A}_{\perp}^{(\tau)}(x,\tau)
			\right]^2
		}.
		\label{eq:transverse-fraction}
	\end{equation}
	
	Finally, the zero of the local scale deformation,
	\begin{equation}
		\partial_\tau p(x_p,\tau)=0,
		\label{eq:pivot-definition}
	\end{equation}
	defines the pivot $x_p$, while the zeros of the transverse geodesic
	acceleration,
	\begin{equation}
		\mathcal{A}_{\perp}^{(\tau)}
		(x_{\kappa,i},\tau)
		=
		0,
		\qquad
		i=1,2,
		\label{eq:curvature-zero-definition}
	\end{equation}
	define the characteristic momentum fractions
	$x_{\kappa,1}$ and $x_{\kappa,2}$. These quantities distinguish the
	momentum-fraction structure associated with motion along the
	statistical trajectory from that associated with its transverse
	bending.

	% ============================================================
	% IV. Geometric structure of the DGLAP trajectory
	% ============================================================
	\section{Geometric Structure of the DGLAP Trajectory}
	\label{sec:geometry-results}

	% ------------------------------------------------------------
	\subsection{From PDF redistribution to Fisher motion}
	\label{subsec:x-resolved-motion}
	
\begin{figure}[htbp]
		\centering
		\includegraphics[width=0.98\columnwidth]
		{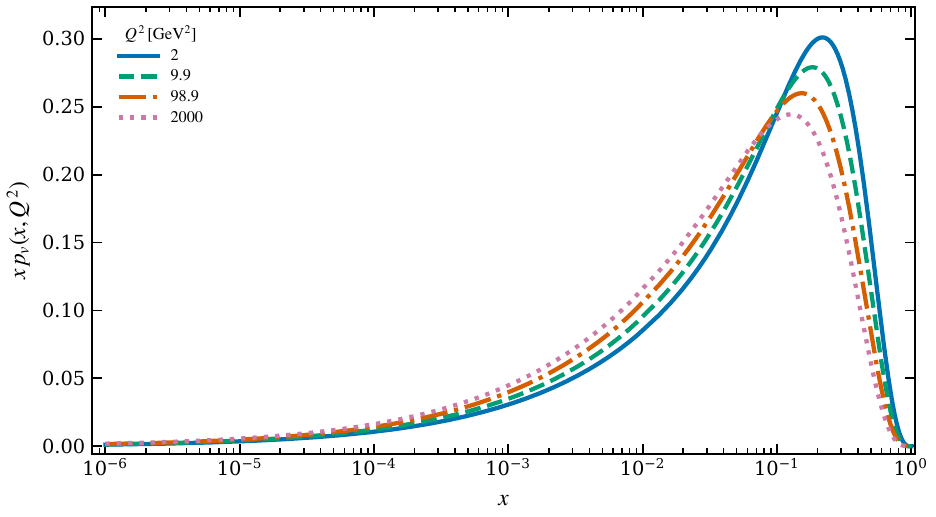}
		\caption{
			Scale evolution of the normalized total-valence distribution,
			displayed as $x\,p(x,Q^2)$ at representative values of $Q^2$.
			Increasing $Q^2$ redistributes the valence density toward lower
			momentum fractions while preserving its normalization. The sequence of
			normalized states defines the Fisher--Rao trajectory studied below.
		}
		\label{fig:pdf-redistribution}
	\end{figure}

	We begin with the underlying redistribution of the normalized valence
	density before examining the geometry of the resulting trajectory.
	Figure~\ref{fig:pdf-redistribution} shows $x\,p(x,Q^2)$ at
	representative scales over
	\begin{equation}
		2~{\rm GeV}^2
		\leq
		Q^2
		\leq
		2000~{\rm GeV}^2 .
	\end{equation}
	
	As the resolution scale increases, the intermediate- and large-$x$
	valence structure is progressively suppressed and the distribution is
	shifted toward lower momentum fractions. Valence-number conservation
	ensures that each scale defines a normalized statistical state along
	the Fisher trajectory. The geometric question is therefore not only
	how the distribution moves in $x$, but how its changing shape
	generates motion in the space of statistical states.
	
\begin{figure*}[htbp]
		\centering
		\includegraphics[width=0.66\textwidth]
		{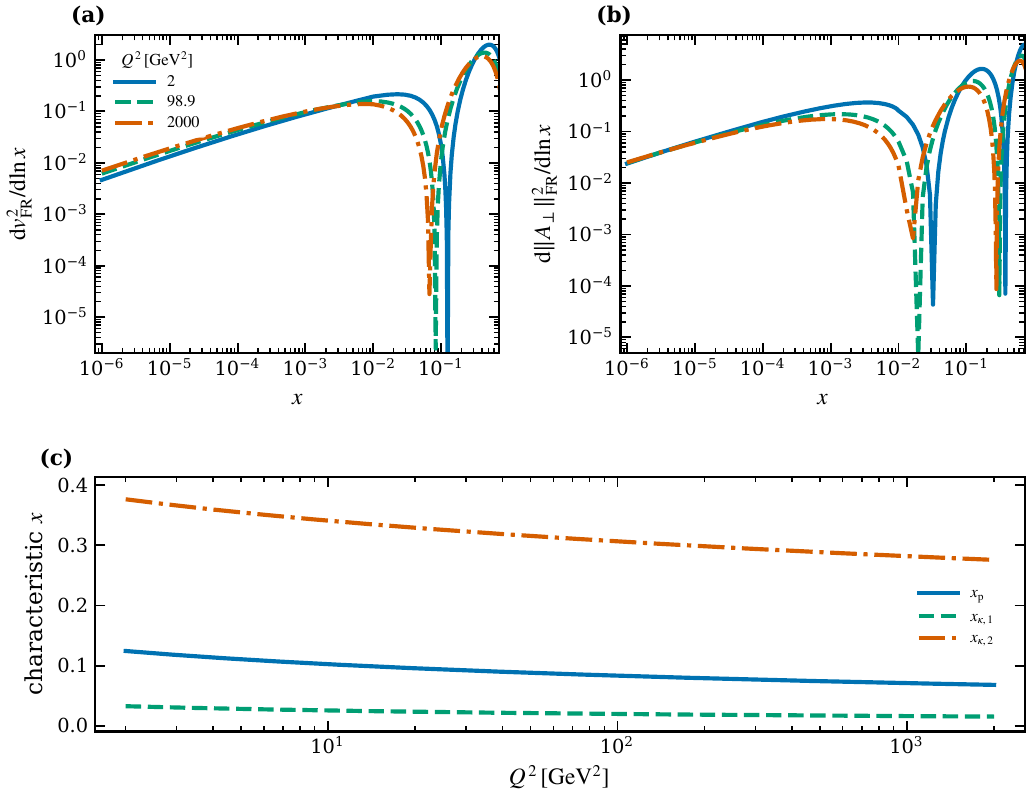}
		\caption{
			Momentum-fraction structure of the LO valence DGLAP trajectory.
			Panel (a) shows the contribution
			$\dd[v_{\rm FR}^{(\tau)}]^2/\dd\ln x$ to the squared Fisher--Rao
			speed. Panel (b) shows the corresponding transverse geodesic-acceleration
			contribution
			$\dd\|\mathcal{A}_{\perp}^{(\tau)}\|^2/\dd\ln x$.
			Panel (c) tracks the pivot $x_p$ and the two transverse-acceleration
			zeros $x_{\kappa,1}$ and $x_{\kappa,2}$. All three characteristic
			momentum fractions migrate toward lower $x$ as $Q^2$ increases, while
			the scales associated with motion along the trajectory and
			transverse bending remain distinct.
		}
		\label{fig:local-information-geometry}
	\end{figure*}

	The local contribution to this motion, per logarithmic interval
	in momentum fraction, is given by the Fisher-speed density in
	Eq.~\eqref{eq:log-fisher-speed-density}.
	Figure~\ref{fig:local-information-geometry}(a) shows that the regions
	contributing most strongly to the Fisher motion need not coincide with
	those carrying the largest probability weight. The Fisher metric is
	sensitive to the relative deformation of the normalized distribution,
	rather than to its local magnitude alone.
	
	A pronounced zero occurs at the scale-dependent pivot $x_p$,
	defined in Eq.~\eqref{eq:pivot-definition}.
	At this momentum fraction the distribution is instantaneously
	unchanged under the scale evolution, while probability is redistributed
	on either side of it. The pivot moves smoothly from
	\begin{equation}
		x_p\simeq0.124
		\qquad
		(Q^2=2~{\rm GeV}^2)
	\end{equation}
	to
	\begin{equation}
		x_p\simeq0.068
		\qquad
		(Q^2=2000~{\rm GeV}^2).
	\end{equation}
	
	The transverse bending of the Fisher trajectory contains a
	different momentum-fraction structure, shown in
	Fig.~\ref{fig:local-information-geometry}(b) and defined by
	Eq.~\eqref{eq:log-transverse-density}.
	Its two characteristic zeros evolve from approximately
	\begin{equation}
		x_{\kappa,1}\simeq0.033,
		\qquad
		x_{\kappa,2}\simeq0.376
	\end{equation}
	at the initial scale to
	\begin{equation}
		x_{\kappa,1}\simeq0.015,
		\qquad
		x_{\kappa,2}\simeq0.276
	\end{equation}
	at the upper endpoint. As summarized in
	Fig.~\ref{fig:local-information-geometry}(c), $x_p$,
	$x_{\kappa,1}$, and $x_{\kappa,2}$ all migrate toward lower $x$ while
	remaining clearly separated. The momentum-fraction structure governing
	motion along the trajectory is therefore distinct from that governing
	its transverse bending.
	
	The same separation appears after integrating over finite
	momentum-fraction regions. At $Q^2=2~{\rm GeV}^2$, the interval
	$0.3<x<0.7$ contains about $14.7\%$ of the normalized probability but
	contributes approximately $51.0\%$ of the squared Fisher speed and
	$21.0\%$ of the transverse-acceleration budget. At
	$Q^2=2000~{\rm GeV}^2$, the same interval contains only about $7.0\%$
	of the probability while still contributing about $40.2\%$ of the
	Fisher-speed budget and $33.1\%$ of the transverse bending.
	Conversely, the region $0.01<x<0.1$ carries about $41.5\%$ of the
	probability at the upper endpoint but contributes only about $7.5\%$
	of the squared Fisher speed.
	
	The Fisher trajectory is therefore governed by the relative
	scale-dependent reshaping of the normalized PDF rather than by the
	location of its largest probability weight. The distinct
	momentum-fraction contributions to statistical displacement and
	transverse bending provide a local picture of the evolution. We next
	examine how the overall rate of Fisher motion changes with scale,
	before turning to the intrinsic curvature and global geometry of
	the trajectory.
	
	% ------------------------------------------------------------
	\subsection{Rate of traversal along the Fisher trajectory}
	\label{subsec:fisher-speed}
	
	Unlike the intrinsic geometry of a trajectory, its
	instantaneous Fisher speed depends on the parameter used
	to describe the evolution.
	
	In the $\tau$ parametrization, the squared Fisher speed
	decreases smoothly from approximately $2.70$ at
	$Q^2=2~{\rm GeV}^2$ to $2.04$ at
	$Q^2=2000~{\rm GeV}^2$, corresponding to a reduction
	of about $24\%$.
	
	To characterize this decrease, we define
	\begin{equation}
		\mathcal{D}_{\tau}
		=
		-\frac{\dd}{\dd\tau}
		\left[v_{\rm FR}^{(\tau)}\right]^2.
		\label{eq:fisher-speed-attenuation}
	\end{equation}
	For the reference trajectory, $\mathcal{D}_{\tau}$
	remains positive throughout the evolution interval
	and decreases from approximately $3.60$ to $2.13$.
	
	\begin{figure*}[htbp]
		\centering
		\includegraphics[width=0.66\textwidth]
		{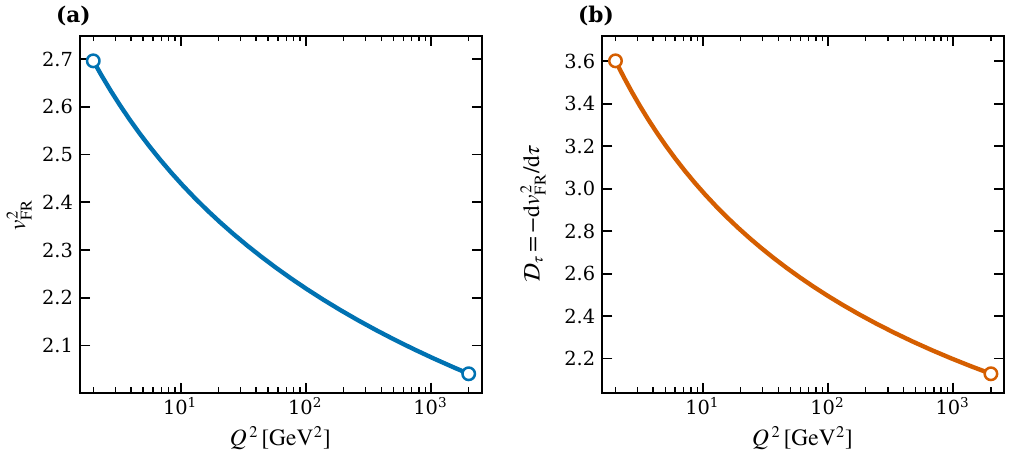}
		\caption{
			Rate of traversal of the LO valence DGLAP trajectory
			in the evolution parameter $\tau$.
			Panel (a) shows the squared Fisher speed
			$\left[v_{\rm FR}^{(\tau)}\right]^2$,
			which decreases smoothly with increasing resolution
			scale. Panel (b) shows the corresponding attenuation
			measure
			$\mathcal{D}_{\tau}
			=-\dd[v_{\rm FR}^{(\tau)}]^2/\dd\tau$.
			Both quantities characterize the rate of statistical
			motion in the $\tau$ parametrization, rather than
			the intrinsic geometric shape of the trajectory.
		}
		\label{fig:fisher-speed-dissipation}
	\end{figure*}
	
	As shown in Fig.~\ref{fig:fisher-speed-dissipation},
	the decreasing Fisher speed indicates that a further
	increment of $\tau$ produces a progressively smaller
	statistical displacement along the reference trajectory.
	This behavior reflects the changing response of the
	normalized valence distribution to the evolution
	operator. It does not, by itself, determine how the
	trajectory bends within the Fisher manifold.
	That distinction is addressed next through the
	geodesic curvature.
	
	% ------------------------------------------------------------
	\subsection{Nongeodesicity and regular geometric bending}
	\label{subsec:nongeodesic-bending}
	
	Having characterized the rate of statistical motion, we turn to the
	intrinsic bending of the Fisher trajectory. The central question is
	whether the DGLAP-selected path follows a Fisher--Rao geodesic.
	The numerical evolution shows a nonzero geodesic curvature throughout
	the analyzed interval. More importantly, this curvature varies only
	weakly with scale, despite the changing rate of traversal established
	in the preceding subsection.
	
\begin{figure*}[htbp]
		\centering
		\includegraphics[width=0.66\textwidth]
		{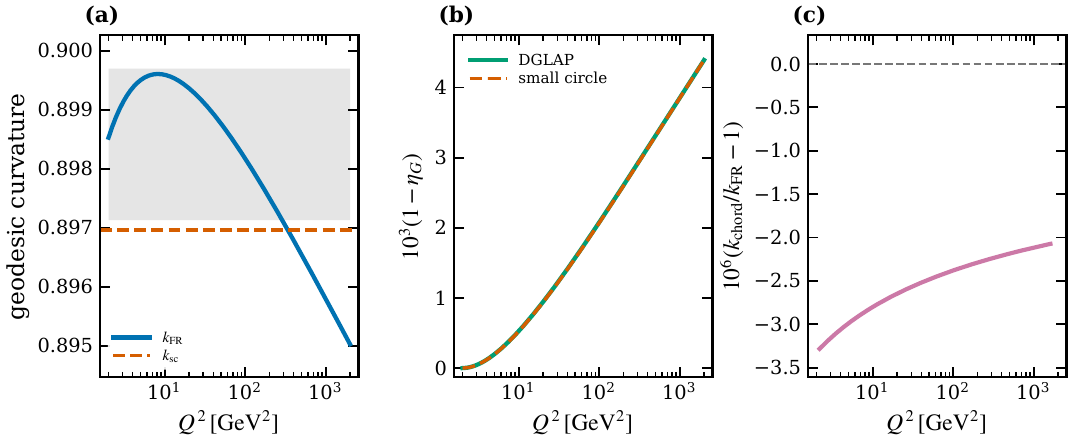}
		\caption{
			Local and global geometric tests for the LO valence DGLAP trajectory.
			Panel (a) compares the dynamically reconstructed geodesic curvature
			$k_{\rm FR}$ with the constant small-circle prediction $k_{\rm sc}$;
			the gray band denotes the arc-length-weighted standard-deviation range
			around $\langle k_{\rm FR}\rangle_s$. Panel (b) compares $10^3(1-\eta_G)$ for the DGLAP trajectory with
			$10^3(1-\eta_G^{\rm sc})$ for the corresponding small circle. Panel (c) shows the relative
			difference
			$10^6(k_{\rm chord}/k_{\rm FR}-1)$ between the finite-chord and
			differential curvature reconstructions.
		}
		\label{fig:geometric-closure}
	\end{figure*}

	Figure~\ref{fig:geometric-closure}(a) shows the curvature reconstructed
	directly from the DGLAP-generated velocity and acceleration. Its Fisher--Rao arc-length average is
	\begin{equation}
		\left\langle k_{\rm FR}\right\rangle_s
		\simeq
		0.8984,
		\label{eq:kfr-mean-result}
	\end{equation}
	with a relative variation of order $10^{-3}$. The departure from
	geodesic motion is therefore highly regular rather than an arbitrary
	bending of the trajectory.
	
	The same trajectory remains close to the metric shortest path in
	an integrated sense. The geodesic length ratio defined in
	Eq.~\eqref{eq:geodesic-length-ratio}, shown in
	Fig.~\ref{fig:geometric-closure}(b), reaches
	\begin{equation}
		\eta_G\simeq0.9956
		\label{eq:eta-result}
	\end{equation}
	at the upper endpoint. The physical DGLAP path therefore exceeds the endpoint geodesic
	distance by only about $0.44\%$ in relative length. Its local curvature is nevertheless clearly nonzero,
	showing that a small excess in total path length can coexist with a
	well-resolved geometric bending.
	
	Panel (c) shows the finite-chord curvature comparison described
	in Sec.~\ref{subsec:numerical-validation}. Its few-parts-per-million
	agreement with the differential reconstruction provides a
	complementary geometric cross-check.
	The nonzero and nearly scale-independent curvature is therefore a
	stable property of the reference trajectory. We next show that its
	magnitude and global path geometry are naturally accounted for by the
	same affine small-circle construction.
	
	% ------------------------------------------------------------
	\subsection{Emergent small-circle closure}
	\label{subsec:small-circle-closure}
	
	The effective dimensionality provides the link between the observed
	curvature and a simple global geometry. The first two principal
	directions capture a fraction
	\begin{equation}
		f_2
		=
		0.99999849
		\label{eq:two-pc-result}
	\end{equation}
	of the arc-length-weighted trajectory variance. Equivalently,
	Eq.~\eqref{eq:plane-residual-pca} gives
	\begin{equation}
		\epsilon_{\rm plane}
		\simeq
		1.23\times10^{-3}.
		\label{eq:plane-residual-result}
	\end{equation}
	The evolution is therefore confined to a narrow neighborhood of a
	two-dimensional affine plane.
	
	The position of this plane relative to the Fisher-sphere center fixes
	the corresponding sphere--plane intersection. For the reference
	trajectory, the plane offset is $h\simeq1.7469$, which gives
	\begin{equation}
		r_{\rm plane}
		=
		\sqrt{R_{\rm FR}^2-h^2}
		\simeq
		0.973796
		\label{eq:radius-plane-result}
	\end{equation}
	and
	\begin{equation}
		k_{\rm sc}
		=
		\frac{|h|}
		{R_{\rm FR}r_{\rm plane}}
		\simeq
		0.8970.
		\label{eq:ksc-result}
	\end{equation}
	The predicted curvature differs from the arc-averaged DGLAP value in
	Eq.~\eqref{eq:kfr-mean-result} by only about
	$1.6\times10^{-3}$ in relative terms. Thus the local nongeodesic
	bending and the global position of the dominant affine plane point to
	the same small-circle geometry.
	
\begin{figure*}[htbp]
		\centering
		\includegraphics[width=0.66\textwidth]
		{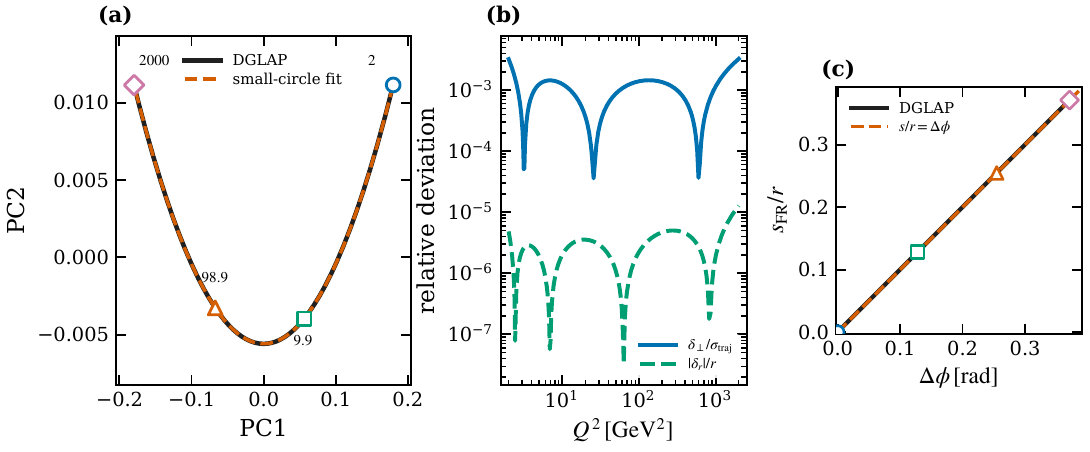}
		\caption{
			Direct small-circle reconstruction of the LO valence DGLAP trajectory.
			Panel (a) shows the trajectory in the dominant two-dimensional affine
			plane together with the circular reconstruction and representative
			evolution scales. Panel (b) compares the normalized out-of-plane
			displacement $\delta_{\perp}/\sigma_{\rm traj}$ with the in-plane
			radial deviation $|\delta r|/r$. Panel (c) compares the accumulated
			Fisher--Rao length $s_{\rm FR}/r$ with the angular displacement
			$\Delta\varphi$; an exact circle satisfies
			$s/r=\Delta\varphi$.
		}
		\label{fig:small-circle-direct}
	\end{figure*}

	Figure~\ref{fig:small-circle-direct}(a) tests this construction directly
	in the dominant plane. The projected DGLAP states trace a smooth arc
	with best-fit radius $r_{\rm fit}\simeq0.973795$, in agreement with
	$r_{\rm plane}\simeq0.973796$ at the $10^{-6}$ relative level.
	
	The two residual measures exhibit a clear hierarchy. Compared
	with the plane residual in Eq.~\eqref{eq:plane-residual-result},
	the in-plane radial residual is
	\begin{equation}
		\epsilon_{\rm rad}
		\simeq
		3.29\times10^{-6}.
		\label{eq:residual-hierarchy}
	\end{equation}
	The dominant departure from the idealized construction is associated
	with the weak nonplanarity of the full trajectory. In their respective
	normalized measures, the residual in-plane noncircularity is more than
	two orders of magnitude smaller than the out-of-plane deviation.
	
	The same closure is visible in the accumulated path length.
	As shown in Fig.~\ref{fig:small-circle-direct}(c), the DGLAP
	trajectory closely follows the exact-circle relation in
	Eq.~\eqref{eq:small-circle-length} throughout the evolution interval. At the
	upper endpoint, the angular span is
	$\Delta\varphi\simeq21.27^\circ$, while
	\begin{equation}
		L_{\rm DGLAP}
		=
		0.361533,
		\qquad
		L_{\rm sc}
		=
		0.361529.
		\label{eq:length-comparison-result}
	\end{equation}
	Their relative difference is
	\begin{equation}
		\frac{
			L_{\rm sc}-L_{\rm DGLAP}
		}{
			L_{\rm DGLAP}
		}
		\simeq
		-1.1\times10^{-5}.
		\label{eq:small-circle-length-closure}
	\end{equation}
	
	The radius, local curvature, radial closure, and accumulated path
	length therefore provide mutually consistent tests of the same
	geometry. Over the finite scale interval considered here, the LO
	DGLAP trajectory follows a short arc close to the small circle selected
	by its dominant affine plane.
	
	% ------------------------------------------------------------
	\subsection{Global Fisher-sphere geometry}
	\label{subsec:global-fisher-geometry}
	
\begin{figure*}[htbp]
		\centering
		\includegraphics[width=0.66\textwidth]
		{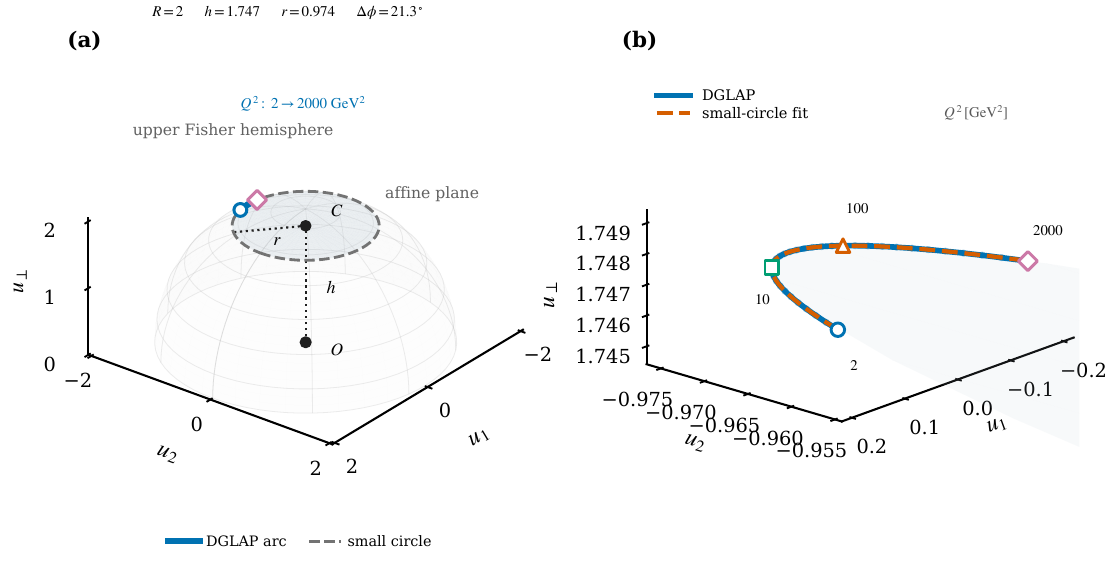}
		\caption{
			Adapted representation of the Fisher-sphere geometry of the LO valence
			DGLAP trajectory. Panel (a) shows the $R_{\rm FR}=2$ Fisher sphere,
			the dominant affine plane, the sphere center $O$, the closest point
			$C$, the offset $h$, and the corresponding small-circle radius $r$.
			The DGLAP evolution lies close to a finite arc of the
			sphere--plane intersection. Panel (b) magnifies the numerical trajectory in the
			adapted coordinates $(u_1,u_2,u_\perp)$, where the ideal affine plane
			lies at $u_\perp=h$. This representation displays the leading
			low-dimensional geometry of the trajectory rather than the full
			infinite-dimensional Fisher embedding space.
		}
		\label{fig:three-dimensional-fisher}
	\end{figure*}

	Figure~\ref{fig:three-dimensional-fisher} summarizes the resulting
	geometry in coordinates adapted to the trajectory. The directions
	$u_1$ and $u_2$ span the dominant affine plane, while $u_\perp$ is
	chosen along the direction from the Fisher-sphere center $O$ to the
	closest point $C$ of that plane. The ideal affine plane is therefore located at $u_\perp=h$.
	
	The geometric parameters obtained from the small-circle
	reconstruction determine the sphere--plane configuration shown in
	Fig.~\ref{fig:three-dimensional-fisher}(a). Since the dominant plane
	does not pass through the sphere center, its intersection with the
	Fisher sphere is a small circle rather than a great circle.
	
	Panel (b) resolves the remaining departure from the ideal affine
	geometry. The dominant evolution occurs along $u_1$ and $u_2$, while
	the displacement in the orthogonal direction remains much smaller.
	With the residuals normalized to the Fisher-sphere radius as
	defined in Appendix~\ref{app:three-dimensional-reconstruction},
	the arc-length-weighted RMS and maximum reconstruction errors are
	$\epsilon_{\rm 3D}^{\rm RMS}\simeq6.4\times10^{-5}$ and
	$\epsilon_{\rm 3D}^{\rm max}\simeq1.7\times10^{-4}$,
	respectively.

	The three-dimensional reconstruction provides
	a geometric visualization of the near-small-circle
	structure identified from the local curvature and
	affine-plane analysis. We next examine how this
	organization is related to the underlying DGLAP
	evolution dynamics.

	% ============================================================
	% V. Dynamical interpretation
	% ============================================================
	
	\section{Dynamical Interpretation}
	\label{sec:dynamical-origin}
	
	The preceding analysis reveals two related features of the
	LO nonsinglet DGLAP trajectory: its effective confinement
	to a low-dimensional affine sector and its nonzero, nearly
	constant geodesic curvature. We examine these observations
	from two complementary perspectives. The Mellin-space
	evolution characterizes the spectral organization of the
	changing parton distribution, while the Fisher-embedded
	acceleration connects the local DGLAP dynamics to the
	geometry of the trajectory.
	
	% ------------------------------------------------------------
	\subsection{Spectral perspective on effective dimensional reduction}
	\label{subsec:mellin-origin}
	
	At LO, the nonsinglet Mellin moments evolve independently according
	to Eq.~\eqref{eq:mellin-solution}, with the anomalous dimensions
	$\gamma_N^{(0)}$ determined by the splitting kernel
	~\cite{BlumleinKurth:1999,Vogt:2005}.
	The conserved mode $\gamma_1^{(0)}=0$ fixes the valence
	normalization, whereas the higher moments describe the
	scale-dependent reshaping of the distribution.
	
	The effective two-dimensional geometry found in
	Sec.~\ref{sec:geometry-results} does not imply a
	two-mode truncation of this Mellin hierarchy. Each
	nontrivial moment retains its own evolution, and the
	full distribution contains contributions from infinitely
	many moments. The observed low dimensionality instead
	characterizes the collective trajectory sampled over
	the finite evolution interval.
	
	This distinction can be made explicit through the
	fixed-operator form of LO evolution,
	\begin{equation}
		\begin{aligned}
			p(\tau)
			&=
			e^{\tau\mathcal{L}}p_0
			\\
			&=
			p_0+\tau\mathcal{L}p_0
			+\frac{\tau^2}{2}\mathcal{L}^2p_0
			+\frac{\tau^3}{3!}\mathcal{L}^3p_0
			+\cdots .
		\end{aligned}
		\label{eq:krylov-expansion}
	\end{equation}
	The successive terms belong to the Krylov sequence
	generated by $\mathcal{L}$ and the initial distribution
	$p_0$. Although this sequence is not generally
	finite dimensional, its contributions to the evolving
	distribution may be strongly correlated over a
	restricted interval in $\tau$.
	
	The Fisher representation introduces the nonlinear
	transformation $\Phi=2\sqrt{p}$. Consequently, the
	principal directions extracted from the Fisher
	trajectory are not individual Mellin eigenmodes.
	They describe collective variations of the normalized
	distribution after the square-root embedding.
	
	The strong two-component dominance observed numerically
	is therefore a finite-window property of the complete
	Fisher trajectory rather than a reduction of the
	underlying DGLAP equation to two independent degrees
	of freedom. The Mellin and Krylov representations
	provide a dynamical perspective on this effective
	compression, while the nearly circular form of the
	trajectory requires additional information about its
	acceleration.
	
	% ------------------------------------------------------------
	\subsection{Geometric organization of the DGLAP acceleration}
	\label{subsec:small-circle-origin}
	
	The geodesic curvature measures how the direction of
	statistical motion changes along the Fisher trajectory.
	For an arc-length parameter $s$, the acceleration on
	the Fisher sphere satisfies
	\begin{equation}
		\frac{\dd^2\Phi}{\dd s^2}
		=
		-\frac{\Phi}{R_{\rm FR}^2}
		+
		k_{\rm FR}N_{\rm FR},
		\label{eq:arc-acceleration-decomposition}
	\end{equation}
	where $N_{\rm FR}$ is the unit direction of the
	geodesic-curvature component. The first term is fixed
	by the spherical normalization constraint, while the
	second describes the intrinsic bending of the path
	away from geodesic motion.
	
	For an exact small circle with center $C$ and radius
	$r$, the acceleration instead takes the form
	\begin{equation}
		\frac{\dd^2\Phi}{\dd s^2}
		=
		-\frac{\Phi-C}{r^2},
		\label{eq:small-circle-acceleration}
	\end{equation}
	with the constant geodesic curvature
	$k_{\rm sc}=|h|/(R_{\rm FR}r)$ from
	Eq.~\eqref{eq:small-circle-curvature}.
	An exact small circle thus combines a fixed affine
	plane with a constant intrinsic curvature.
	
	The numerical DGLAP trajectory displays a close
	realization of this geometry. Its dominant affine
	plane is displaced from the Fisher-sphere center,
	while the geodesic curvature reconstructed directly
	from the evolution-generated acceleration varies
	only weakly with scale. The agreement between
	$k_{\rm FR}$ and the corresponding small-circle
	value $k_{\rm sc}$ connects the local bending of the
	physical trajectory to its global affine geometry.
	
	This correspondence is particularly informative
	because spherical normalization alone does not
	require either near-planarity or nearly constant
	geodesic curvature. These properties characterize
	the particular path selected by the LO evolution
	and its initial distribution.
	
	The connection to DGLAP dynamics follows from
	Eqs.~\eqref{eq:dglap-tau} and~\eqref{eq:second-tau-derivative}.
	The velocity and acceleration of the Fisher state
	are therefore determined by successive actions
	of the same nonsinglet evolution operator, together
	with the nonlinear square-root embedding. In this
	representation, the geodesic curvature characterizes
	the changing direction of PDF reshaping, rather than
	the magnitude of the redistribution alone.
	
	The $x$-resolved results provide a complementary
	partonic picture. Different momentum-fraction regions
	contribute to the Fisher speed and to the transverse
	geodesic acceleration, showing that the regions
	driving statistical displacement need not be those
	primarily responsible for its bending.
	
	The distinction between the shape of the path and
	its rate of traversal is also important. The running
	coupling determines the mapping from $Q^2$ to $\tau$
	and contributes to the physical-scale dependence
	of the Fisher speed. Even in the $\tau$
	parametrization, however, the speed changes as the
	normalized distribution evolves. By contrast, the
	geodesic curvature is invariant under monotonic
	reparameterization. The decreasing Fisher speed
	and the nearly constant curvature therefore describe
	different aspects of the same DGLAP trajectory.
	
	Taken together, the local curvature and global
	small-circle reconstruction indicate that the
	scale-dependent reshaping of the reference valence
	distribution follows a highly organized geometric
	path. The approximately constant bending is an
	observed feature of this evolution, consistent with
	the low-dimensional structure of its Fisher
	representation.
	
	% ------------------------------------------------------------
	\subsection{Physical scope and further directions}
	\label{subsec:scope-universality}
	
	The Fisher-sphere construction follows generally
	from the normalization of a positive probability
	density. The near-small-circle structure, however,
	is a property of the particular evolution trajectory
	rather than a consequence of normalization alone.
	The present results concern the GRV98-based total-valence
	distribution evolved with the LO nonsinglet kernel over
	$2\leq Q^2\leq2000~{\rm GeV}^2$.
	
	Within the fixed-operator LO formulation, the
	running coupling controls the relation between
	the physical scale and the evolution parameter
	$\tau$. For a fixed initial distribution, this
	reparameterization leaves the geometric path
	unchanged. By contrast, changing the initial
	valence distribution changes the trajectory
	itself. Separate evolution of $u_v$ and $d_v$,
	or alternative valence inputs, can therefore
	probe the dependence of the observed geometry
	on the initial partonic structure.
	
	Further extensions involve singlet--gluon
	evolution, where the coupled matrix kernel
	introduces additional dynamical components
	~\cite{FurmanskiPetronzio:1980}, and higher-order
	splitting functions
	~\cite{MochVermaserenVogt:2004,
		VogtMochVermaseren:2004}.
	Modern global PDF determinations provide
	additional reference distributions for such
	comparisons~\cite{Hou:2021,Bailey:2021,NNPDF:2021}.
	
	These developments offer a systematic way to
	examine how the information geometry changes
	with the initial state and the evolution
	dynamics. More broadly, the present construction
	demonstrates that Fisher--Rao geometry can
	resolve both the local bending and the global
	organization of a QCD evolution trajectory,
	complementing its conventional description
	through PDFs and Mellin moments.

	% ============================================================
	% VI. Summary and conclusions
	% ============================================================
	
	\section{Summary and Conclusions}
	\label{sec:conclusions}
	
	We have investigated the information geometry of LO nonsinglet
	DGLAP evolution by treating normalized valence distributions
	as points on a Fisher--Rao statistical manifold. Through the
	isometric embedding $\Phi=2\sqrt{p}$, the scale evolution
	defines a trajectory on a Hilbert sphere of radius two.
	The DGLAP-generated velocity and acceleration provide a
	direct measure of its geodesic curvature, allowing the
	geometry of the QCD-selected path to be examined using
	the standard LO nonsinglet DGLAP kernel.
	
	For the GRV98-based total-valence trajectory over
	$2\leq Q^2\leq2000~{\rm GeV}^2$, we find that the evolution
	is not a Fisher--Rao geodesic, but exhibits a nonzero,
	weakly scale-dependent geodesic curvature. More than
	$99.999\%$ of the trajectory variance is captured by
	two affine directions. Their associated plane is displaced
	from the Fisher-sphere center and defines a small-circle
	reference geometry. The curvature predicted by this
	construction closely agrees with that obtained directly
	from the DGLAP-generated acceleration, while the projected
	trajectory and accumulated path length are consistent
	with the corresponding circular geometry. These results
	characterize the reference evolution as a highly regular,
	near-small-circle trajectory over the finite scale interval.
	
	The $x$-resolved analysis further shows that Fisher motion
	reflects the relative reshaping of the normalized PDF,
	rather than its probability weight alone. Different
	momentum-fraction regions contribute to the statistical
	displacement and transverse bending of the trajectory,
	providing a partonic interpretation of its local geometry.
	The Mellin representation offers a complementary dynamical
	perspective: the effective low-dimensional Fisher trajectory
	is associated with the collective evolution of the full
	spectral hierarchy, without requiring a truncation to a
	small number of moments.
	
	The present study thus establishes a quantitative framework
	for examining the geometry of QCD scale evolution and
	identifies an organized near-small-circle structure in a
	representative LO nonsinglet trajectory. Its dependence
	on the initial parton distribution and evolution dynamics
	can be explored through alternative valence inputs,
	singlet--gluon mixing, and higher perturbative orders.
	More generally, the Fisher--Rao description complements
	the conventional analysis of PDFs and Mellin moments by
	characterizing the geometric path selected by QCD evolution.

	% ============================================================
	% Acknowledgments
	% ============================================================
	\begin{acknowledgments}
		This work was supported by the National Natural Science Foundation of China
		(Grant Nos. 12547118 and 12605167).
	\end{acknowledgments}

	% ============================================================
	% APPENDICES
	% ============================================================
	
	\appendix
	
	% ============================================================
	% Appendix A
	% ============================================================
	
	\section{Reparameterization Invariance}
	\label{app:parameterization}
	
	We briefly establish the parameterization invariance of
	the geodesic curvature used in Sec.~\ref{sec:formalism}.
	
	Consider two monotonic parameters $\lambda$ and
	$\widetilde{\lambda}$ related by
	\begin{equation}
		a=\frac{\dd\widetilde{\lambda}}{\dd\lambda}>0.
	\end{equation}
	The derivatives of the Fisher state transform as
	\begin{align}
		\partial_\lambda\Phi
		&=a\,\partial_{\widetilde{\lambda}}\Phi,
		\\
		\partial_\lambda^2\Phi
		&=
		a^2\partial_{\widetilde{\lambda}}^2\Phi
		+\frac{\dd a}{\dd\lambda}
		\partial_{\widetilde{\lambda}}\Phi.
	\end{align}
	
	Substituting these relations into the transverse
	geodesic residual defined in
	Eq.~\eqref{eq:geodesic-residual}, the term proportional
	to $\dd a/\dd\lambda$ cancels under the tangential
	projection. Consequently,
	\begin{equation}
		\mathcal{R}_{\rm geo}^{(\lambda)}
		=
		a^2\mathcal{R}_{\rm geo}^{(\widetilde{\lambda})}.
	\end{equation}
	Since
	$v_\lambda^2=a^2v_{\widetilde{\lambda}}^2$,
	the geodesic curvature and arc-length element satisfy
	\begin{equation}
		\boxed{
			k_{\rm FR}^{(\lambda)}
			=
			k_{\rm FR}^{(\widetilde{\lambda})},
			\qquad
			v_\lambda\,\dd\lambda
			=
			v_{\widetilde{\lambda}}\,
			\dd\widetilde{\lambda}.
		}
	\end{equation}
	
	Thus the transformation from
	$t=\ln(Q^2/Q_0^2)$ to $\tau$ changes the rate
	of traversal but leaves the geometric path,
	its accumulated length, and its geodesic
	curvature unchanged.

	% ============================================================
	% Appendix B
	% ============================================================
	
	\section{Numerical Implementation and Validation}
	\label{app:numerics}
	
	This appendix collects the numerical prescriptions
	used to construct the reference LO nonsinglet
	trajectory and its Fisher--Rao observables.
	
	% ------------------------------------------------------------
	\subsection{Input distributions and evolution settings}
	\label{app:input-running}
	
	At the original GRV98 LO input scale
	$Q_{\rm GRV}^2=0.26~{\rm GeV}^2$, we use
	~\cite{GluckReyaVogt:1998}
	\begin{align}
		x u_v(x,Q_{\rm GRV}^2)
		&=
		1.239\,x^{0.48}(1-x)^{2.72}
		\nonumber\\
		&\quad\times
		\left(1-1.8\sqrt{x}+9.5x\right),
		\\
		d_v(x,Q_{\rm GRV}^2)
		&=
		0.614(1-x)^{0.9}
		u_v(x,Q_{\rm GRV}^2).
	\end{align}
	
	The input is evolved to $Q_0^2=2~{\rm GeV}^2$
	before the geometric analysis. The one-loop
	running coupling follows Eq.~\eqref{eq:one-loop-running},
	with
	\begin{equation}
		\begin{aligned}
			\Lambda_{\rm LO}^{(3)}&=0.204~{\rm GeV},\\
			\Lambda_{\rm LO}^{(4)}&=0.175~{\rm GeV},\\
			\Lambda_{\rm LO}^{(5)}&=0.132~{\rm GeV}.
		\end{aligned}
	\end{equation}
	The active flavor number changes from three
	to four at $Q^2=1.96~{\rm GeV}^2$ and from
	four to five at $Q^2=20.25~{\rm GeV}^2$.
	The full geometric-analysis interval corresponds to
	\begin{equation}
		\Delta\tau=0.2365413.
	\end{equation}
	
	The reference $x$ grid contains $N_x=800$ points.
	Logarithmic spacing is used for
	$10^{-6}\leq x<10^{-2}$, linear spacing for
	$10^{-2}\leq x<0.9$, and logarithmic refinement
	in $1-x$ near the upper endpoint
	$x_{\max}=1-10^{-8}$. These regions contain
	$40\%$, $35\%$, and $25\%$ of the grid points,
	respectively.
	
	Integration on the $x$ grid uses trapezoidal quadrature weights.
	The convolution uses $N_z=128$ Gauss--Legendre nodes and interpolation
	at off-grid arguments $x/z$. The plus-prescription subtraction is
	implemented directly in the convolution, without an explicit endpoint
	regulator. Evolution is integrated with a fourth-order
	Runge--Kutta method using
	$\Delta\tau_{\max}=2.5\times10^{-4}$.
	The resulting trajectory is sampled at 241
	uniformly spaced values of $\tau$.
	
	% ------------------------------------------------------------
	\subsection{Discrete normalization and derivatives}
	\label{app:discrete-evolution}
	
	Let $\widetilde{\mathcal{L}}_{\rm d}$ denote
	the discrete LO convolution operator, and let
	$w_j$ be the quadrature weights on the represented
	$x$ interval. Define
	\begin{equation}
		c[f]
		=
		\sum_j w_j
		\left(\widetilde{\mathcal{L}}_{\rm d}f\right)_j.
	\end{equation}
	For a normalized discrete density,
	$\sum_jw_jp_j=1$, the evolution velocity is
	\begin{equation}
		V[p]
		=
		\widetilde{\mathcal{L}}_{\rm d}p
		-c[p]p.
		\label{eq:app-conservative-velocity}
	\end{equation}
	This prescription enforces
	$\sum_jw_jV_j=0$ on the numerical grid.
	
	Differentiating the same normalized flow gives
	the discrete second derivative
	\begin{equation}
		B[p]
		=
		\widetilde{\mathcal{L}}_{\rm d}V
		-c[V]p-c[p]V.
		\label{eq:app-conservative-acceleration}
	\end{equation}
	The Fisher velocity and acceleration are
	constructed from $V[p]$ and $B[p]$, rather
	than by finite differences of sampled states.
	
	The correction in
	Eq.~\eqref{eq:app-conservative-velocity}
	defines a normalization-preserving discrete
	flow. Its continuum interpretation requires
	distinguishing the normalization on the
	represented interval from that of the full
	distribution on $0<x<1$.
	
	% ------------------------------------------------------------
	\subsection{Discrete Fisher geometry}
	\label{app:discrete-fisher}
	
	The quadrature-weighted Fisher state is
	represented by
	\begin{equation}
		Z_{ij}
		=
		2\sqrt{w_jp(x_j,Q_i^2)},
		\label{eq:app-discrete-fisher-state}
	\end{equation}
	so that ordinary Euclidean inner products of
	the vectors $Z_i$ reproduce the discretized
	Fisher--Rao inner product.
	
	The PCA weights are assigned from the
	accumulated Fisher arc length. Writing
	$\Delta s_i=s_{i+1}-s_i$ and
	$L=\sum_i\Delta s_i$, the trapezoidal weights are
	\begin{equation}
		\begin{aligned}
			\omega_0
			&=\frac{\Delta s_0}{2L},\\
			\omega_i
			&=\frac{\Delta s_{i-1}+\Delta s_i}{2L},
			\\
			\omega_{N-1}
			&=\frac{\Delta s_{N-2}}{2L}.
		\end{aligned}
	\end{equation}
	They satisfy $\sum_i\omega_i=1$ and define
	the weighted covariance operator introduced
	in Eq.~\eqref{eq:weighted-covariance}.
	
	The characteristic momentum fractions
	$x_p$ and $x_{\kappa,i}$ are determined
	from sign changes of the corresponding
	evolution derivatives, with their zeros
	interpolated in $\ln x$.
	
	% ------------------------------------------------------------
	\subsection{Three-dimensional reconstruction}
	\label{app:three-dimensional-reconstruction}
	
	For Fig.~\ref{fig:three-dimensional-fisher},
	we retain the two dominant PCA directions
	$e_1,e_2$ and the normalized plane-offset
	direction $\widehat C=C/\|C\|$.
	
	The reconstructed state is
	\begin{equation}
		\Phi_i^{(3)}
		=
		\sum_{a=1}^{2}
		\langle\Phi_i,e_a\rangle e_a
		+
		\langle\Phi_i,\widehat C\rangle\widehat C.
	\end{equation}
	The reconstruction residuals, normalized
	to the Fisher-sphere radius, are
	\begin{equation}
		\begin{aligned}
			\epsilon_{\rm 3D}^{\rm RMS}
			&=
			\frac{1}{R_{\rm FR}}
			\left[
			\sum_i\omega_i
			\|\Phi_i-\Phi_i^{(3)}\|^2
			\right]^{1/2},
			\\
			\epsilon_{\rm 3D}^{\rm max}
			&=
			\frac{1}{R_{\rm FR}}
			\max_i\|\Phi_i-\Phi_i^{(3)}\|.
		\end{aligned}
	\end{equation}
	The corresponding numerical values are reported in
	Sec.~\ref{subsec:global-fisher-geometry}.
	
	% ------------------------------------------------------------
	
	% ------------------------------------------------------------
	\subsection{Numerical cross-checks}
	\label{app:numerical-checks}
	
	The numerical evolution is benchmarked against the
	analytic LO nonsinglet solution in Mellin space.
	For the splitting-kernel convention used here,
	the anomalous dimensions are
	\begin{equation}
		\gamma_N^{(0)}
		=
		C_F\left[
		\frac{3}{2}-2H_N+\frac{1}{N(N+1)}
		\right],
		\qquad
		H_N=\sum_{j=1}^{N}\frac{1}{j}.
	\end{equation}
	The first five Mellin moments are used in the
	comparison, including the conserved first moment
	$\gamma_1^{(0)}=0$.
	
	The discrete normalization is monitored both
	before and after applying the conservative
	correction in
	Eq.~\eqref{eq:app-conservative-velocity}.
	The corrected flow preserves the discrete
	normalization to machine precision, while the
	uncorrected first-moment drift provides a
	diagnostic of quadrature and interpolation errors.
	
	As a geometric cross-check, the differential
	geodesic curvature is compared with a
	finite-chord estimator constructed from
	neighboring Fisher states. Five sampled states
	on either side of each central point are used
	to define the local arc and its endpoint chord.
	This comparison tests the consistency between
	the operator-generated derivatives and the
	geometry of the sampled trajectory.
	
	The numerical settings and principal validation
	diagnostics are summarized in
	Table~\ref{tab:numerical-validation}.
	
	\begin{table}[tbp]
		\caption{
			Reference numerical settings and validation
			diagnostics for the LO nonsinglet DGLAP trajectory.
		}
		\label{tab:numerical-validation}
		\begin{ruledtabular}
			\begin{tabular}{lc}
				Quantity & Reference value \\
				\hline
				\multicolumn{2}{l}{\textit{Numerical settings}} \\[2pt]
				$N_x$ & $800$ \\
				$N_z$ & $128$ \\
				$x_{\min}$ & $10^{-6}$ \\
				$1-x_{\max}$ & $10^{-8}$ \\
				$\Delta\tau_{\max}$ & $2.5\times10^{-4}$ \\
				Number of sampled states & $241$ \\
				$\Delta\tau$ ($Q^2:2\rightarrow2000~{\rm GeV}^2$)
				& $0.2365413$ \\
				\hline
				\multicolumn{2}{l}{\textit{Validation diagnostics}} \\[2pt]
				Maximum normalization error
				& $2.22\times10^{-16}$ \\
				Minimum sampled density $p(x,Q^2)$
				& $3.04\times10^{-35}$ \\
				Maximum uncorrected first-moment drift
				& $4.88\times10^{-4}$ \\
				Maximum Mellin relative error
				& $3.46\times10^{-4}$ \\
				Maximum finite-chord curvature mismatch
				& $3.28\times10^{-6}$ \\
			\end{tabular}
		\end{ruledtabular}
	\end{table}
	
	These checks establish the internal consistency
	of the reference numerical implementation.
	The reported diagnostics characterize the
	specified discretization and do not constitute
	a systematic continuum-extrapolation study.

	% ============================================================
	% Bibliography
	% ============================================================
	\bibliographystyle{apsrev4-2}
	\bibliography{refs}
	
\end{document}